# Generalized Feedback Control Modeling Method for Control-Driven Converter Systems

Chen Zhang, *Member, IEEE*, Haoxiang Zong, *Member, IEEE*, Xu Cai and Marta Molinas, *Fellow, IEEE*

***Abstract*—Converters-based systems like wind farms manifest themselves as control-intensive systems, where control-driven stability issues frequently occur, e.g., oscillations. Such issues are popularly studied via circuit impedance-based methods. However, given its implicit controller modeling trait, the impedance-based methods have limitations in system analysis and designs involving large-scale controllers. To address this issue, this paper presents a novel frequency domain modeling framework, as a perspective shift from the circuit to the control system. Since the obtained model features a multi-input-multi-output (MIMO) feedback control structure and *explicit controller placement*, it is termed the Generalized Feedback Control (GFC) model. GFC modeling is conducted for both single and multi-converter cases, and the resulting models are validated by frequency scan and stability test. Moreover, advantages of the GFC method in achieving interaction analysis and stability-oriented designs of multi-controllers are demonstrated by three application examples, further suggesting its great potential for being applied to the analysis and design issues of converter systems involving large-scale controllers.**

## I. Introduction

RENEWABLE power generation (RPG) systems, e.g., wind and solar power plants, manifest themselves as converters-dominated and control-intensive systems [1]. Recently, field practices in operating RPGs have shown that they are prone to oscillatory stability issues [2] when connected to weak grids. Such stability issues evidently arising from the interactions among multi-entities (e.g., the RPG and the grid) fall precisely within the research scope of *port-interaction* stability.

To study the above issue, developing the system's port-based models would be the first step. Although admittedly interaction dynamics are mainly driven by the converter's controls, their outcome eventually appears on the circuit system, thus, modeling the system as circuit port-based *impedances* becomes a natural thought. To this aim, various analytical models of grid-tied converters in different domains have been proposed in the past years, e.g., the *dq* impedance [3], the $\alpha\beta$ impedance [4], the sequence impedance [5], and the modified sequence domain impedance [6]. These impedance models are typically in a multi-input and multi-output (MIMO) form, more precisely, a 2-by-2 matrix with non-negligible off-diagonals mainly attributed to the *dq-asymmetry* property of the applied converter controls [7]. Despite of different model formations, they are equivalent in terms of stability judgement if being rigorously modeled [8]. Therefore, by applying either of them, an impedance-based source-load circuit for the grid-tied converter system can be established, where the converter is often defined as the "load" while the grid is the "source" [9]. This circuit system can be further reformulated into a *feedback control* structure, as a result of which stability assessment becomes trivial and can be achieved by evaluating the system's open-loop gain [10] (i.e., the impedance ratio of the source and the load) with the Generalized Nyquist Criterion (GNC) [11].

Further, frequency-domain (FD) analysis of the impedance-depicted converter systems can also lead to various stabilization methods for oscillation suppressions. In general, these methods can be referred to as the *impedance reshaping* since their underlying mechanism of achieving stabilization effects is obtained by altering the impedance's characteristics. Moreover, implementation of this idea is shown to be diverse, e.g., by optimizing controller's parameters [12], [13], adding ancillary damping loops [14]-[16], and performing robust designs [17], [18], etc. The majority of them are single-variable based (e.g., tuning the controller of a single control loop), while existing practices suggests that an oscillation is generally caused by the collective effects of multiple controllers and from different control loops. Thus, to acquire an effective suppression, it is preferred to beforehand perform a sensitivity analysis to find controllers that most dominate the concerned oscillation mode [19]. Despite that, this method is still exposed to a risk of losing its effectiveness when the mode is inevitably sensitive to multi-controllers, e.g., the wind farm grid integration system in [20]. For such cases, a multivariable-oriented method is desired.

In this respect, for instance, [21] presents a multi-controller-parameter tuning method using the impedance model of a grid-tied converter system. The effect of the tuning is to achieve an online tracking of a set of reference closed-loop poles so that the stability of the converter under varying grid conditions can be largely guaranteed. Whereas, tuning the multi-parameters of an ancillary damping loop is conducted in [22] under the state-space (SS) framework and is achieved by solving an SS model-

This paper is supported by National Natural Science Foundation of China (No. 52207215) and grants from the Delta Power Electronics Science and Education Development Program of Delta Group (No. DREK2023004).

Chen Zhang, Haoxiang Zong, and Xu Cai are with the Department of Electrical Engineering, Shanghai Jiao Tong University, Shanghai, China. (e-mails:{nealbc, haoxiangzong, xucai} @sjtu.edu.cn ).

Marta Molinas is with the Department of Engineering Cybernetics, Norwegian University of Science and Technology, Trondheim, Norway. (e-mail: marta.molinas@ntnu.no).

formulated stability optimization problem. In general, the SS- and the FD-framework-based analyses are exchangeable for most occasions, while subtle differences may lie in that: the SS method is usually recognized as a more friendly approach for multivariable control issues [23] due to its explicit modeling of the system's states; while the FD method is known for its compact model structure and wider application range, e.g., the ability to perform grey/black-box analysis [24], which is useful for cases if only frequency responses are available. However, current implementations of the two frameworks may encounter challenges when applied to tuning or design of large-scale controllers, e.g., for a multi-converter system case. In detail, the main challenge for the SS framework lies in its difficulty of effectively accounting for multi-resonance characteristics of transmission systems (mostly modeled as algebraic equations [25], [26]), of which the issue is proven crucial for control interaction analysis. On the other hand, given the implicit controller modeling issue of the impedance-based FD method, its deficiency in the evaluation and design of large-scale controllers is also evident. Therefore, an effective and efficient model adequately competent for control-intensive system analysis and design remains under exploration.

To this end, this paper will present a novel FD modeling framework and method for converters-dominated systems, referred to as the Generalized Feedback Control (GFC) method. Merits of this method mainly lie in the following aspects:

1) The GFC model for either single- or multi-converter systems holds the structure of a MIMO feedback control system, more importantly, controller *clustering* and controller *explicit* placement are achieved. This controller-explicit model paves the way for easier control-driven analysis and designs.

2) The GFC model is an FD model, thus merits of existing FD models, e.g., modularity and the ability of performing frequency response-based analysis, are all inherited.

3) Existing FD tools can be used as well, particularly, the FD-Modal-Analysis (FDMA) method [27]—a powerful tool for extracting FD eigen-sensitivities of a given system's model

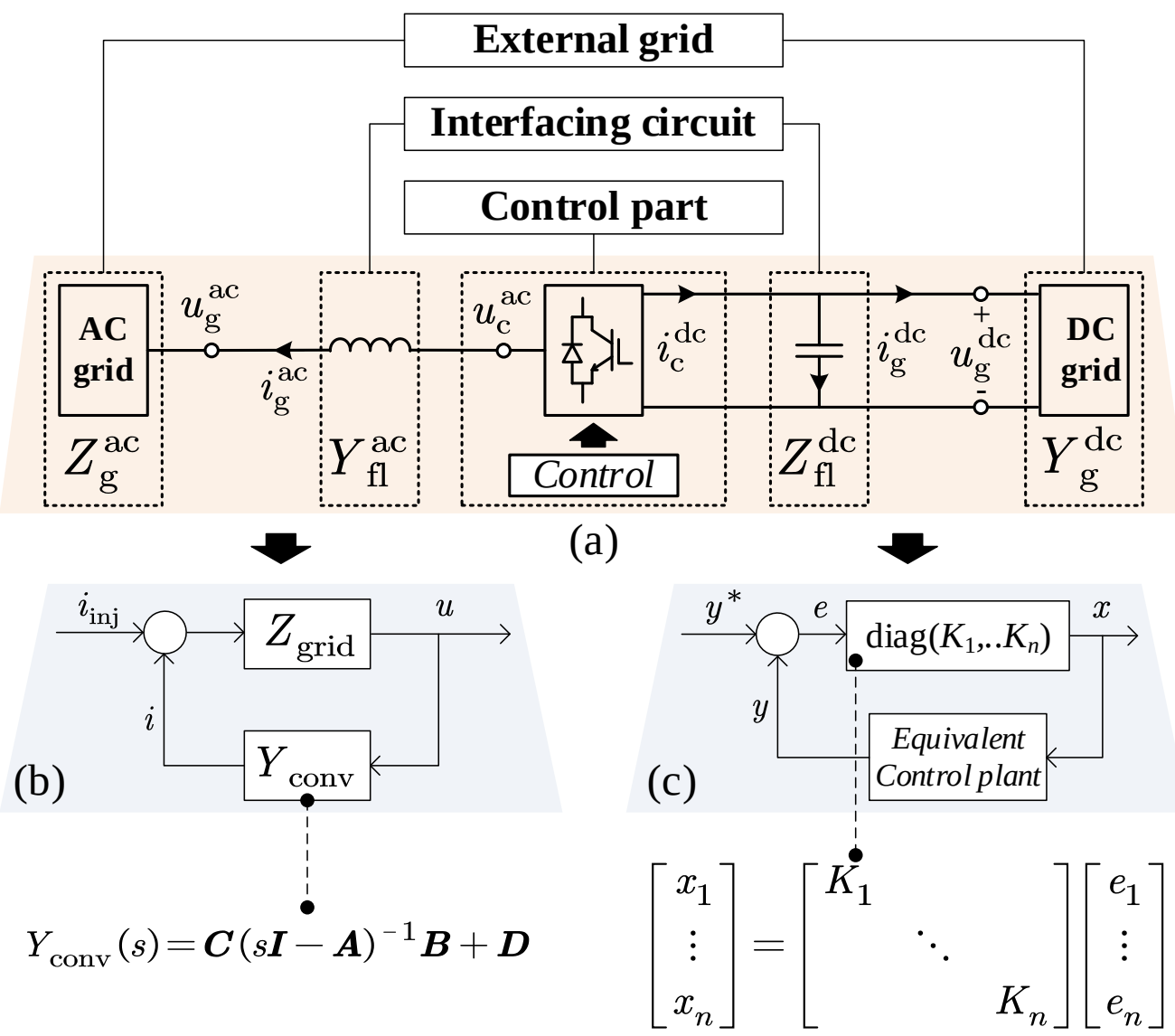

Fig. 1 (a) A typical grid-tied converter system. (b) Impedance-based control diagram; (c) GFC-based control diagram, where $K_n$ denotes one of controllers inside the converter's control scheme; "diag" denotes a diagonal matrix.

(e.g., nodal admittance model of a multi-converter system) [28] and thus often being used for oscillation diagnostics and source locating [29]. However, the FDMA at present is commonly performed with impedance models, of which the sensitivities of controllers are difficult to acquire due to the implicit controller modeling issue, while the GFC model is exempt from this issue due to its merit of explicit controller placement. The rest of the paper is organized as follows:

Section II. aims to show the general idea and framework of the GFC modeling method, due to which details of converters are omitted. Section III. presents the GFC modeling process of both single and multi-converter cases along with their model validations. To demonstrate the superiority of the GFC method, examples of multi-controller interaction analysis and stability-oriented reshaping designs are provided in Section IV. Finally, conclusions of the paper are drawn in Section V.

## II. GFC Modeling Framework

### A. Overview

The way to analyze a MIMO linear system (1) can be diverse since different choices of input $\boldsymbol{u}$ and output $\boldsymbol{y}$ can lead to different modeling and analysis view-angles. For example, the impedance-based analysis is achieved if the inputs/outputs for modeling are defined as individual device's circuit ports.

$$\begin{aligned} \dot{\boldsymbol{x}} &= \boldsymbol{A}\boldsymbol{x} + \boldsymbol{B}\boldsymbol{u} \\ \boldsymbol{y} &= \boldsymbol{C}\boldsymbol{x} + \boldsymbol{D}\boldsymbol{u} \end{aligned} \tag{1}$$

Moreover, with impedances, although the grid-tied converter system in Fig. 1(a) can be formulated into a feedback control structure as in Fig. 1(b), the controllers are implicitly inside the system's model. By contrast, under the GFC framework as depicted in Fig. 1(c), a new feedback control system but with explicit controller placement (can be from different control loops and different converters) will be obtained. Apparently, the GFC model is expected to be more friendly for control-driven analysis and design.

To intuitively achieve the GFC modeling, the overall system in Fig. 1(a) is split into three subsystems, i.e., 1) the ac/dc side interfacing circuit; 2) converter's internal controls, and 3) the external ac/dc grids. In which, the interfacing circuit is where the external grid interacts with the converter's internal control. As further explained in the Fig. 2, the ac side output ($u_{\mathrm{c}}^{\mathrm{ac}}$) of the converter can interact with the ac grid's output ($u_{\mathrm{g}}^{\mathrm{ac}}$) via the ac side filter ($Y_{\mathrm{fl}}^{\mathrm{ac}}$), and the resulting current response ($i_{\mathrm{g}}^{\mathrm{ac}}$) will flow back to the ac grid. On the other hand, the interaction between the dc side output ($i_{\mathrm{c}}^{\mathrm{dc}}$) of converter and the dc grid's output ($i_{\mathrm{g}}^{\mathrm{dc}}$) is accomplished via the dc side filter ($Z_{\mathrm{fl}}^{\mathrm{dc}}$), of which the response ($u_{\mathrm{g}}^{\mathrm{dc}}$) will flow back to the dc grid. In general, the introduction of the interfacing circuit layer can better result in a modularized modeling process, particularly for multi-converter systems, this will be shown later.

Next, the GFC modeling framework for a single converter case will be firstly introduced, then followed by its extension to the case of multi-converter systems.

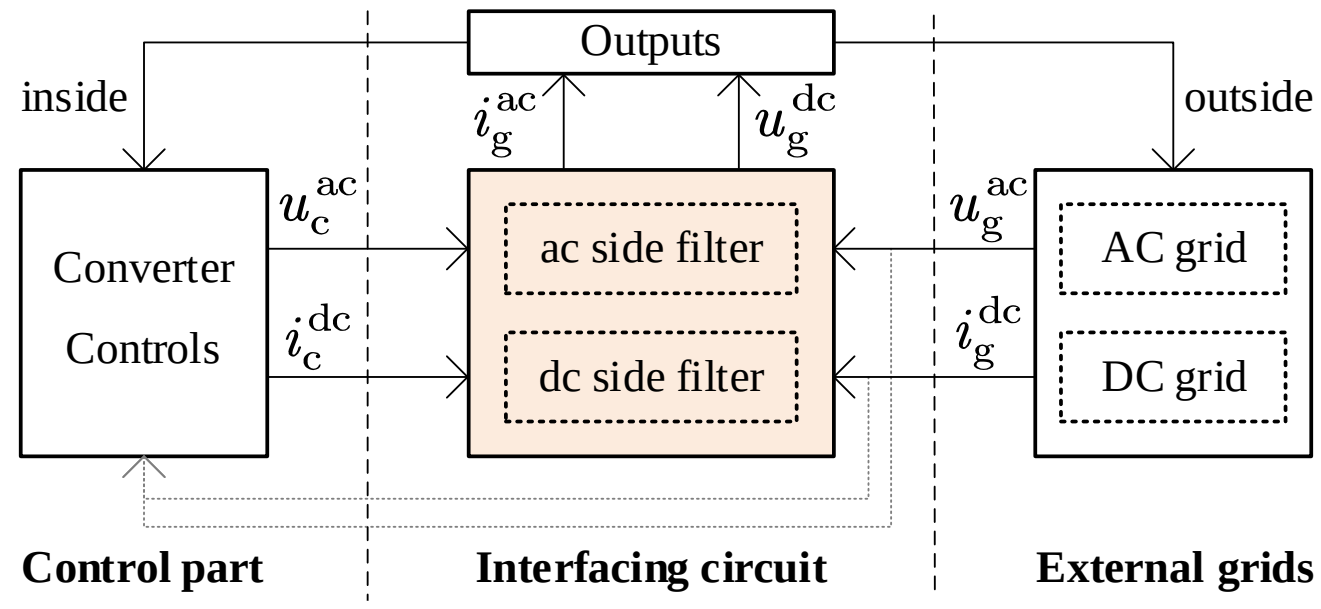

Fig. 2 Depiction of the connection and interaction of each part.

*B. GFC Modeling Framework for Single Converter Case*

*1) GFC Modeling of Converter Control System*

The linearized control diagram of converters is typically composed of the *building-block* as in Fig. 3(a), i.e., a summator with its net output connected to a transfer function. For better illustration, $K_i$ denotes *i*-th controller with its input and output as $e_i$ and $x_i$. $r_i$ can either denote a constant reference (e.g., $y_{\mathrm{i}}^{\mathrm{ref}}$) if the controller is at the outermost loop or another controller's output (e.g., $x_k$). $y_i$ denotes the feedback input, and lastly $u_{\mathrm{ext},i}$ denotes the external input, e.g., for converters, it usually refers to the outputs of ac and dc grids, i.e., $u_{\mathrm{g}}^{\mathrm{ac}}$, $i_{\mathrm{g}}^{\mathrm{dc}}$.

To better achieve GFC modeling, the traditional summator is generalized to a *linear mapping operator* $L\{\}$ as:

$$L\{\boldsymbol{x}\} = \sum a_i x_i \tag{2}$$

where $\boldsymbol{x}$ denotes a set, $x_i$ is the element in that set, $a_i$ is the linear mapping. Notably, $x_i$ can be vectors and the mapping $a_i$ can be matrices (with elements either be constants or *s*-domain transfer-functions), which dependents on the specific cases.

With this operator, $e_{\mathrm{c},i}$ in Fig. 3(b) can be written as linear combinations of vectors $\boldsymbol{x}_{\mathrm{fl}}$, $\boldsymbol{x}_{\mathrm{c}}$, $\boldsymbol{u}_{\mathrm{ext}}$, where the external input vector $\boldsymbol{u}_{\mathrm{ext}}$ mainly contains $u_{\mathrm{g}}^{\mathrm{ac}}$, $i_{\mathrm{g}}^{\mathrm{dc}}$. Whereas $\boldsymbol{x}_{\mathrm{c}}$, $\boldsymbol{x}_{\mathrm{fl}}$ denotes outputs of the controllers and the interfacing circuit (i.e., $i_{\mathrm{g}}^{\mathrm{ac}}$, $u_{\mathrm{g}}^{\mathrm{dc}}$), respectively. Based on this, the *i*-th generalized building block can be formulated as:

$$\begin{aligned} &\text{forward:} \quad K_i e_{\mathrm{c},i} = x_{\mathrm{c},i} \\ &\text{feedback:} \quad e_{\mathrm{c},i} = L\{\boldsymbol{x}_{\mathrm{c}}, \boldsymbol{x}_{\mathrm{fl}}, \boldsymbol{u}_{\mathrm{ext}}\} \end{aligned} \tag{3}$$

By collecting all the considered controller blocks, the forward and feedback relations can be obtained as in (4), which is in fact the GFC model for the control part.

$$\begin{aligned} &\text{forward:} \quad \underbrace{\begin{bmatrix} x_{\mathrm{c},1} \\ \vdots \\ x_{\mathrm{c},n} \end{bmatrix}}_{\boldsymbol{x}_{\mathrm{c}}} = \underbrace{\begin{bmatrix} K_1 & & \\ & \ddots & \\ & & K_n \end{bmatrix}}_{\boldsymbol{K}} \underbrace{\begin{bmatrix} e_{\mathrm{c},1} \\ \vdots \\ e_{\mathrm{c},n} \end{bmatrix}}_{\boldsymbol{e}_{\mathrm{c}}} \\ &\text{feedback:} \quad \boldsymbol{e}_{\mathrm{c}} = L\{\boldsymbol{x}_{\mathrm{c}}, \boldsymbol{x}_{\mathrm{fl}}, \boldsymbol{u}_{\mathrm{ext}}\} \end{aligned} \tag{4}$$

*2) GFC Modeling of Interfacing Circuit*

According to Fig. 1(a), the model for the interfacing circuit can be directly written with basic circuit laws as:

$$\begin{bmatrix} i_{\mathrm{g}}^{\mathrm{ac}} \\ u_{\mathrm{g}}^{\mathrm{dc}} \end{bmatrix} = \underbrace{\begin{bmatrix} Y_{\mathrm{fl}}^{\mathrm{ac}} & \\ & Z_{\mathrm{fl}}^{\mathrm{dc}} \end{bmatrix}}_{\boldsymbol{G}_{\mathrm{fl}}} \begin{bmatrix} u_{\mathrm{c}}^{\mathrm{ac}} - u_{\mathrm{g}}^{\mathrm{ac}} \\ i_{\mathrm{c}}^{\mathrm{dc}} - i_{\mathrm{g}}^{\mathrm{dc}} \end{bmatrix} \tag{5}$$

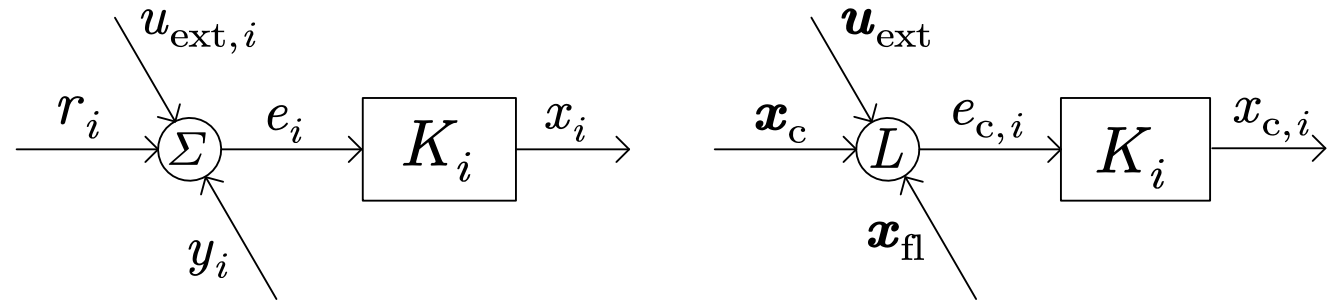
Fig. 3 The building block of a linearized control system. (a) traditional block definition; (b) generalized block definition for GFC modeling.

Interestingly, this model resembles the forward gain model of a controller block under the GFC framework, where $Y_{\mathrm{fl}}^{\mathrm{ac}}$ and $Z_{\mathrm{fl}}^{\mathrm{dc}}$ can be regarded as associated gains. Given this observation, the interfacing circuits can be modeled as "*virtual controllers*", thus being unified with the GFC model of controllers. To better achieve this, the below notations are used for consistency:

$$\boldsymbol{x}_{\mathrm{fl}} = \begin{bmatrix} x_{\mathrm{fl}}^{\mathrm{ac}} \\ x_{\mathrm{fl}}^{\mathrm{dc}} \end{bmatrix} = \begin{bmatrix} i_{\mathrm{g}}^{\mathrm{ac}} \\ u_{\mathrm{g}}^{\mathrm{dc}} \end{bmatrix}, \text{ and } \boldsymbol{e}_{\mathrm{fl}} = \begin{bmatrix} u_{\mathrm{c}}^{\mathrm{ac}} - u_{\mathrm{g}}^{\mathrm{ac}} \\ i_{\mathrm{c}}^{\mathrm{dc}} - i_{\mathrm{g}}^{\mathrm{dc}} \end{bmatrix} = \begin{bmatrix} e_{\mathrm{fl}}^{\mathrm{ac}} \\ e_{\mathrm{fl}}^{\mathrm{dc}} \end{bmatrix} \tag{6}$$

Then, the forward relation of the virtual controller can be written as in (7), while its feedback relation is developed next.

$$\text{forward:} \quad \boldsymbol{G}_{\mathrm{fl}} \boldsymbol{e}_{\mathrm{fl}} = \boldsymbol{x}_{\mathrm{fl}} \tag{7}$$

First, it is noticed that for typical controls, $u_{\mathrm{c}}^{\mathrm{ac}}$ (i.e., converter ac side output as earlier defined) usually refers to the output of innermost control loop and can be represented as linear combinations of elements from $\boldsymbol{x}_{\mathrm{fl}}$ and $\boldsymbol{x}_{\mathrm{c}}$. On the other hand, by applying the ac-dc side power balance of converter, $i_{\mathrm{c}}^{\mathrm{dc}}$ (i.e., converter dc side output) can also be found linearly dependent on $\boldsymbol{x}_{\mathrm{fl}}$ and $\boldsymbol{x}_{\mathrm{c}}$. Thus, the outputs of converter can be written as:

$$\begin{bmatrix} u_{\mathrm{c}}^{\mathrm{ac}} & i_{\mathrm{c}}^{\mathrm{dc}} \end{bmatrix}^{\mathrm{T}} = L\{\boldsymbol{x}_{\mathrm{c}}, \boldsymbol{x}_{\mathrm{fl}}\} \tag{8}$$

Then, according to the definition of $\boldsymbol{e}_{\mathrm{fl}}$ in (6), it has:

$$\text{feedback:} \quad \boldsymbol{e}_{\mathrm{fl}} = L\{\boldsymbol{x}_{\mathrm{c}}, \boldsymbol{x}_{\mathrm{fl}}\} - \boldsymbol{u}_{\mathrm{ext}} = L\{\boldsymbol{x}_{\mathrm{c}}, \boldsymbol{x}_{\mathrm{fl}}, \boldsymbol{u}_{\mathrm{ext}}\} \tag{9}$$

Therefore, (7) and (9) result in the GFC model for the interfacing circuit or the virtual controller.

Besides, from (4) and (9) it can be seen that both feedbacks contain the external inputs from the grids. To find the model in a closed-form, the GFC model of external grid should be known, which is shown next.

*3) GFC Modeling of AC and DC-side Grids*

GFC modeling of the ac/dc grid is identical to its impedance modeling. For a single converter case, the grid is with a simple form, which is usually represented by a Thevenin equivalent circuit, and its model can be directly written as:

$$\boldsymbol{u}_{\mathrm{ext}} = \begin{bmatrix} u_{\mathrm{g}}^{\mathrm{ac}} \\ i_{\mathrm{g}}^{\mathrm{dc}} \end{bmatrix} = \underbrace{\begin{bmatrix} Z_{\mathrm{g}}^{\mathrm{ac}} & \\ & Y_{\mathrm{g}}^{\mathrm{dc}} \end{bmatrix}}_{\boldsymbol{G}_{\mathrm{grid}}} \begin{bmatrix} i_{\mathrm{g}}^{\mathrm{ac}} \\ u_{\mathrm{g}}^{\mathrm{dc}} \end{bmatrix} = \boldsymbol{G}_{\mathrm{grid}} \boldsymbol{x}_{\mathrm{fl}} \tag{10}$$

where $Z_{\mathrm{g}}^{\mathrm{ac}}$ and $Y_{\mathrm{g}}^{\mathrm{dc}}$ denote the ac grid impedance and dc grid admittance, respectively.

*4) The Final GFC Model*

To find the final GFC model, the above three parts' models will be assembled. First, given the ac/dc interfacing circuits are modeled as virtual controllers, their forward gains can be merged with those of the control part, this can be achieved by appending (7) to (4), resulting in the final forward gain $\boldsymbol{G}(s)$ of the overall system as:

$$\begin{bmatrix} x_{\mathrm{c},1} \\ \vdots \\ x_{\mathrm{c},n} \\ x_{\mathrm{fl}}^{\mathrm{ac}} \\ x_{\mathrm{fl}}^{\mathrm{dc}} \end{bmatrix} = \underbrace{\begin{bmatrix} K_1 & & & & \\ & \ddots & & & \\ & & K_n & & \\ & & & Y_{\mathrm{fl}}^{\mathrm{ac}} & \\ & & & & Z_{\mathrm{fl}}^{\mathrm{dc}} \end{bmatrix}}_{\boldsymbol{G}(s)} \begin{bmatrix} e_{\mathrm{c},1} \\ \vdots \\ e_{\mathrm{c},n} \\ e_{\mathrm{fl}}^{\mathrm{ac}} \\ e_{\mathrm{fl}}^{\mathrm{dc}} \end{bmatrix} \tag{11}$$

The above equation can be compactly written as:

$$\text{forward:} \quad \begin{bmatrix} \boldsymbol{x}_{\mathrm{c}} \\ \boldsymbol{x}_{\mathrm{fl}} \end{bmatrix} = \begin{bmatrix} \boldsymbol{K} & \\ & \boldsymbol{G}_{\mathrm{fl}} \end{bmatrix} \begin{bmatrix} \boldsymbol{e}_{\mathrm{c}} \\ \boldsymbol{e}_{\mathrm{fl}} \end{bmatrix} \tag{12}$$

Along with the merging of the forward relations, the corresponding feedback relations (i.e., the control part (4) and the virtual controller part (9)) can be combined as:

$$\begin{bmatrix} \boldsymbol{e}_{\mathrm{c}} \\ \boldsymbol{e}_{\mathrm{fl}} \end{bmatrix} = \begin{bmatrix} \bar{\mathbf{A}}_{\mathrm{c}} & \bar{\mathbf{B}}_{\mathrm{c}} & \bar{\mathbf{C}}_{\mathrm{c}} \\ \bar{\mathbf{A}}_{\mathrm{fl}} & \bar{\mathbf{B}}_{\mathrm{fl}} & \bar{\mathbf{C}}_{\mathrm{fl}} \end{bmatrix} \begin{bmatrix} \boldsymbol{x}_{\mathrm{c}} \\ \boldsymbol{x}_{\mathrm{fl}} \\ \boldsymbol{u}_{\mathrm{ext}} \end{bmatrix} \tag{13}$$

where $\bar{\mathbf{A}}_{\mathrm{c/fl}}$, $\bar{\mathbf{B}}_{\mathrm{c/fl}}$, $\bar{\mathbf{C}}_{\mathrm{c/fl}}$ are mapping matrices for $\boldsymbol{e}_{\mathrm{c}}$ and $\boldsymbol{e}_{\mathrm{fl}}$. It can be seen that the feedback relation contains the external input from the grids. To find the model in its closed form, $\boldsymbol{u}_{\mathrm{ext}}$ should be eliminated by substituting (10) into (13), which yields the final feedback gain $\boldsymbol{F}(s)$ of the overall system as:

$$\begin{bmatrix} \boldsymbol{e}_{\mathrm{c}} \\ \boldsymbol{e}_{\mathrm{fl}} \end{bmatrix} = \underbrace{\begin{bmatrix} \bar{\mathbf{A}}_{\mathrm{c}} & \bar{\mathbf{B}}_{\mathrm{c}} + \bar{\mathbf{C}}_{\mathrm{c}} \boldsymbol{G}_{\mathrm{grid}} \\ \bar{\mathbf{A}}_{\mathrm{fl}} & \bar{\mathbf{B}}_{\mathrm{fl}} + \bar{\mathbf{C}}_{\mathrm{fl}} \boldsymbol{G}_{\mathrm{grid}} \end{bmatrix}}_{\boldsymbol{F}(s)} \begin{bmatrix} \boldsymbol{x}_{\mathrm{c}} \\ \boldsymbol{x}_{\mathrm{fl}} \end{bmatrix} \tag{14}$$

Finally, the overall system's GFC model is obtained in (12) and (14). From this modeling framework introduction, it can also be seen that modeling $\boldsymbol{G}(s)$ almost takes no effort, while the feedback matrix $\boldsymbol{F}(s)$ seems not. However, as will be shown later, the acquisition of $\boldsymbol{F}(s)$ is also overall simple.

## C. GFC Modeling Framework for Multi-Converter Systems

Once the GFC model for the single converter is established, as depicted in Fig. 4, derivation of a multi-converter system's (with homogenous control structure) GFC model is almost a two-step procedure: the first step is to vectorize model elements of the single converter per the number of multi-converters; the second is to assemble the dimension-extended partial models

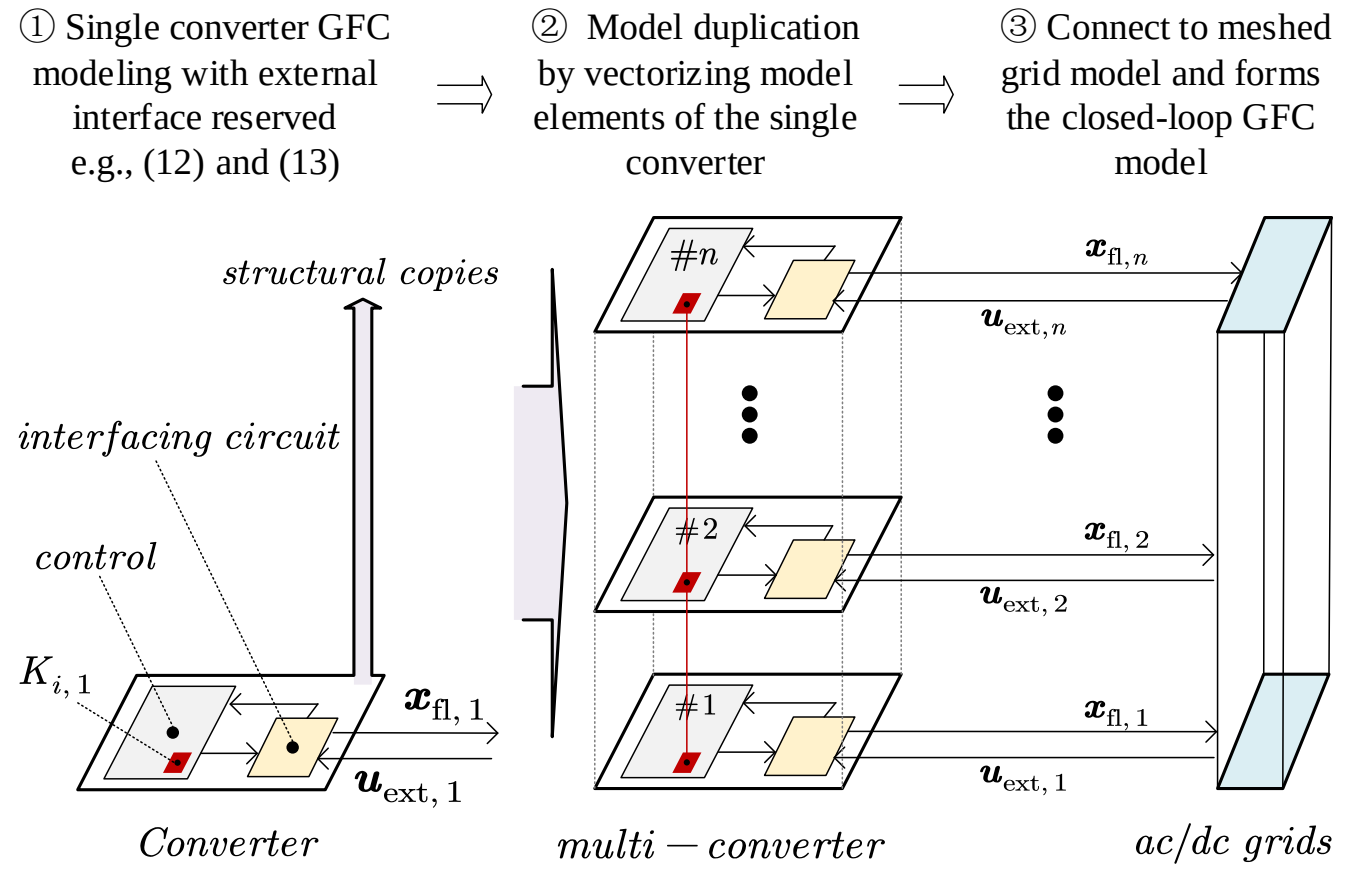

Fig. 4 Depiction on the GFC modeling of a multi-converter system.

with the model of the external grids to obtain the final GFC model in its closed-form.

### 1) Modeling of the Meshed AC and DC Grid

The grid model for a single converter case in (10) is simple, while for the multi-converter system, the number of interfacing nodes to the ac and dc grids will increase according to the number of converters. This issue can be accounted for by vectorizing the elements in (10) as:

$$[\boldsymbol{u}]_{\mathrm{ext}} = \begin{bmatrix} [u]_{\mathrm{g}}^{\mathrm{ac}} \\ [i]_{\mathrm{g}}^{\mathrm{dc}} \end{bmatrix} = \underbrace{\begin{bmatrix} \boldsymbol{Z}_{\mathrm{net}}^{\mathrm{ac}} & \\ & \boldsymbol{Y}_{\mathrm{net}}^{\mathrm{dc}} \end{bmatrix}}_{\boldsymbol{G}_{\mathrm{net}}} \begin{bmatrix} [i]_{\mathrm{g}}^{\mathrm{ac}} \\ [u]_{\mathrm{g}}^{\mathrm{dc}} \end{bmatrix} = \mathbf{G}_{\mathrm{net}} [\boldsymbol{x}]_{\mathrm{fl}} \tag{15}$$

where the operator '$[x]$' denotes expanding the dimension of the variable $x$ by $n$ (i.e., vectorization), and $n$ is the number of nodes or converters, e.g., $[u]_{\mathrm{g}}^{\mathrm{ac}} = \left[u[1]_{\mathrm{g}}^{\mathrm{ac}}, \ldots, u[n]_{\mathrm{g}}^{\mathrm{ac}}\right]^{\mathrm{T}}$. Worth noting here that in practice the converter is usually modeled in *dq-frame*. In this case, when the operator is applied, it applies to the intact *dq*-block rather than its individual *d*- or *q*-channels.

On the other hand, $\boldsymbol{Z}_{\mathrm{net}}^{\mathrm{ac}}$ and $\boldsymbol{Y}_{\mathrm{net}}^{\mathrm{dc}}$ denote the models of meshed ac and dc grids. They can all be obtained by the nodal admittance construction approach (please see [30] for details), since the method is well established, the process is omitted here.

### 2) Modeling of Interfacing Circuit for Multi-Converters

The forward gain of the interfacing circuit can be obtained similarly as the meshed grid by applying the vectorization, as:

$$[\boldsymbol{x}]_{\mathrm{fl}} = \begin{bmatrix} [x]_{\mathrm{g}}^{\mathrm{ac}} \\ [x]_{\mathrm{g}}^{\mathrm{dc}} \end{bmatrix} = \underbrace{\begin{bmatrix} [Y]_{\mathrm{fl}}^{\mathrm{ac}} & \\ & [Z]_{\mathrm{fl}}^{\mathrm{dc}} \end{bmatrix}}_{[\boldsymbol{G}]_{\mathrm{fl}}} \begin{bmatrix} [e]_{\mathrm{fl}}^{\mathrm{ac}} \\ [e]_{\mathrm{fl}}^{\mathrm{dc}} \end{bmatrix} = [\boldsymbol{G}]_{\mathrm{fl}} [e]_{\mathrm{fl}} \tag{16}$$

where $[Y]_{\mathrm{fl}}^{\mathrm{ac}} = \mathrm{diag}\left(Y[1]_{\mathrm{fl}}^{\mathrm{ac}}, \ldots, Y[n]_{\mathrm{fl}}^{\mathrm{ac}}\right)$ is a diagonal matrix as the filters of converters are independent from each other; the vectorized dc side filter $[Z]_{\mathrm{fl}}^{\mathrm{dc}}$ is in a similar form.

Next, based on the feedback model $\boldsymbol{e}_{\mathrm{fl}}$ of the single converter case in (13), the counterparts for multi-converter case can be obtained by vectorizing all the elements in $\bar{\mathbf{A}}_{\mathrm{fl}}$, $\bar{\mathbf{B}}_{\mathrm{fl}}$, $\bar{\mathbf{C}}_{\mathrm{fl}}$ as:

$$[\boldsymbol{e}]_{\mathrm{fl}} = [\bar{\mathbf{A}}]_{\mathrm{fl}} [\boldsymbol{x}]_{\mathrm{c}} + [\bar{\mathbf{B}}]_{\mathrm{fl}} [\boldsymbol{x}]_{\mathrm{fl}} + [\bar{\mathbf{C}}]_{\mathrm{fl}} [\boldsymbol{u}]_{\mathrm{ext}} \tag{17}$$

where, e.g., vectorizing $a_{ij}$ in $\bar{\mathbf{A}}_{\mathrm{fl}}$ has the formation:

$$[a_{ij}]_{\mathrm{fl}} = \begin{bmatrix} a_{ij}[1] & & \\ & \ddots & \\ & & a_{ij}[n] \end{bmatrix} \tag{18}$$

### 3) GFC model of the multi-converter system

First, the final forward gain of the multi-converter system can be obtained by simply vectorizing (12) as:

$$\begin{bmatrix} [x]_{\mathrm{c},1} \\ \vdots \\ [x]_{\mathrm{c},n} \\ [\boldsymbol{x}]_{\mathrm{fl}}^{\mathrm{ac}} \end{bmatrix} = \begin{bmatrix} [K]_1 & & & \\ & \ddots & & \\ & & [K]_n & \\ & & & [\boldsymbol{G}]_{\mathrm{fl}}^{\mathrm{ac}} \end{bmatrix} \begin{bmatrix} [e]_{\mathrm{c},1} \\ \vdots \\ [e]_{\mathrm{c},n} \\ [\boldsymbol{e}]_{\mathrm{fl}}^{\mathrm{ac}} \end{bmatrix} \tag{19}$$

where $[K]_j = \mathrm{diag}(K[1]_j, \ldots, K[n]_j)$ i.e., collecting the same $K_j$ controller from $n$ converters. Correspondingly, the vectorized feedback relation of (13) for multi-converter system can be derived similarly as (17), thus omitted here. To further find the feedback relation in its closed form, the external inputs should be eliminated by using the models of meshed ac/dc grids in (15). Thereby, the final feedback model of the multi-converter system is obtained as:

$$\begin{bmatrix} [\boldsymbol{e}]_{\mathrm{c}} \\ [\boldsymbol{e}]_{\mathrm{fl}} \end{bmatrix} = \begin{bmatrix} [\bar{\mathbf{A}}]_{\mathrm{c}} & [\bar{\mathbf{B}}]_{\mathrm{c}} + [\bar{\mathbf{C}}]_{\mathrm{c}} \boldsymbol{G}_{\mathrm{net}} \\ [\bar{\mathbf{A}}]_{\mathrm{fl}} & [\bar{\mathbf{B}}]_{\mathrm{fl}} + [\bar{\mathbf{C}}]_{\mathrm{fl}} \boldsymbol{G}_{\mathrm{net}} \end{bmatrix} \begin{bmatrix} [\boldsymbol{x}]_{\mathrm{c}} \\ [\boldsymbol{x}]_{\mathrm{fl}} \end{bmatrix} \tag{20}$$

Finally, (19) and (20) make up the GFC model framework for the multi-converter system.

It is noted from the multi-converter system's GFC model that not only the explicit controller placement is obtained but also the clustering of controllers according to their types is naturally achieved thanks to the vectorization method. This model structure merit enables the easier interaction analysis of different types of controllers under multi-converters cases.

## III. GFC Modeling Examples And Validation

In this section, detailed GFC modeling for both a single converter and a multi-converter system will be provided, along with model validations via frequency scan and stability tests.

### A. GFC Method Applied to the Single Converter Case

#### 1) GFC Modeling Details

A typical grid-tied converter system is considered in Fig. 5(a).

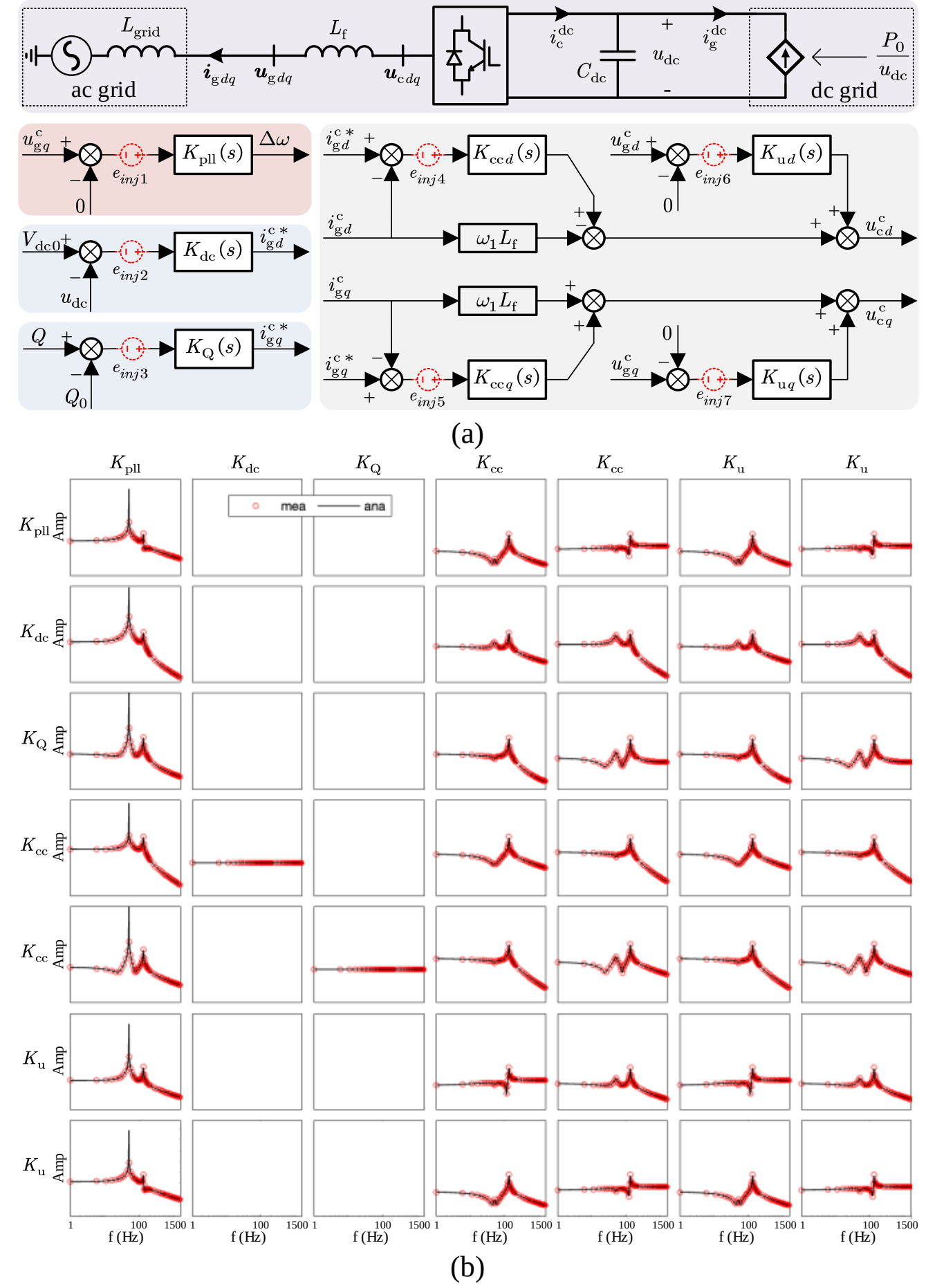

Fig. 5. GFC model validation for a single converter system. (a) control scheme; (b) comparison of analytical model and frequency scanning results.

In which, relevant controllers are: the phase-locked-loop (PLL) $K_{\mathrm{pll}}(s)$, dc voltage controller $K_{\mathrm{dc}}(s)$, reactive power controller $K_{\mathrm{Q}}(s)$, current controllers $K_{\mathrm{cc}d}(s)$, $K_{\mathrm{cc}q}(s)$, and the ac grid voltage feedforward controller $K_{\mathrm{u}d}(s)$, $K_{\mathrm{u}q}(s)$. The *dq* impedance for the ac filter is $\begin{bmatrix} sL_{\mathrm{fl}} & -\omega_1 L_{\mathrm{fl}} \\ \omega_1 L_{\mathrm{fl}} & sL_{\mathrm{fl}} \end{bmatrix}$, while its admittance $L_{\mathrm{fl}}$ is denoted by $\begin{bmatrix} Y^{\mathrm{ac}}_{\mathrm{fl}dd}(s) & Y^{\mathrm{ac}}_{\mathrm{fl}dq}(s) \\ Y^{\mathrm{ac}}_{\mathrm{fl}qd}(s) & Y^{\mathrm{ac}}_{\mathrm{fl}qq}(s) \end{bmatrix}$. The dc-side filter impedance $Z^{\mathrm{dc}}_{\mathrm{fl}}(s)$ for this case is $[sC_{\mathrm{dc}}]^{-1}$. Based on (4), the following forward matrix $\boldsymbol{G}(s)$ can be directly obtained:

$$\begin{bmatrix} x_{\mathrm{c},1} \\ x_{\mathrm{c},2} \\ x_{\mathrm{c},3} \\ x_{\mathrm{c},4} \\ x_{\mathrm{c},5} \\ x_{\mathrm{c},6} \\ x_{\mathrm{c},7} \\ x^{\mathrm{ac}}_{\mathrm{fl}d} \\ x^{\mathrm{ac}}_{\mathrm{fl}q} \\ x^{\mathrm{dc}}_{\mathrm{fl}} \end{bmatrix} = \underbrace{\begin{bmatrix} K_{\mathrm{pll}} &&&&&&&&& \\ & K_{\mathrm{dc}} &&&&&&&& \\ && K_{\mathrm{Q}} &&&&&&& \\ &&& K_{\mathrm{cc}d} &&&&&& \\ &&&& K_{\mathrm{cc}q} &&&&& \\ &&&&& K_{\mathrm{u}d} &&&& \\ &&&&&& K_{\mathrm{u}q} &&& \\ &&&&&&& Y^{\mathrm{ac}}_{\mathrm{fl}dd} & Y^{\mathrm{ac}}_{\mathrm{fl}dq} & \\ &&&&&&& Y^{\mathrm{ac}}_{\mathrm{fl}qd} & Y^{\mathrm{ac}}_{\mathrm{fl}qq} & \\ &&&&&&&&& Z^{\mathrm{dc}}_{\mathrm{fl}} \end{bmatrix}}_{\boldsymbol{G}(s)} \begin{bmatrix} e_{\mathrm{c},1} \\ e_{\mathrm{c},2} \\ e_{\mathrm{c},3} \\ e_{\mathrm{c},4} \\ e_{\mathrm{c},5} \\ e_{\mathrm{c},6} \\ e_{\mathrm{c},7} \\ e^{\mathrm{ac}}_{\mathrm{fl}d} \\ e^{\mathrm{ac}}_{\mathrm{fl}q} \\ e^{\mathrm{dc}}_{\mathrm{fl}} \end{bmatrix} \tag{21}$$

It is worth of mentioning that, $\boldsymbol{G}(s)$ can be transformed into a strictly diagonal matrix if *dq* elements are symmetric [7] and the modified sequence domain transformation is applied [6].

Next, the output variables $\boldsymbol{x}_{\mathrm{c}}$ and $\boldsymbol{x}_{\mathrm{fl}}$ can be instantiated by the actual variables in Fig. 5(a), as:

$$\begin{aligned} &x_{\mathrm{c},1} = \Delta\omega, \quad x_{\mathrm{c},2} = \Delta i^{\mathrm{c}*}_{\mathrm{g}d}, \quad x_{\mathrm{c},3} = \Delta i^{\mathrm{c}*}_{\mathrm{g}q} \\ &x_{\mathrm{c},4} = \Delta u^{\mathrm{c}}_{\mathrm{c}d1}, \ x_{\mathrm{c},5} = \Delta u^{\mathrm{c}}_{\mathrm{c}q1}, \ x_{\mathrm{c},6} = \Delta u^{\mathrm{c}}_{\mathrm{c}d2}, \ x_{\mathrm{c},7} = \Delta u^{\mathrm{c}}_{\mathrm{c}q2} \\ &x^{\mathrm{ac}}_{\mathrm{fl}d} = \Delta i^{\mathrm{ac}}_{\mathrm{g}d}, \quad x^{\mathrm{ac}}_{\mathrm{fl}q} = \Delta i^{\mathrm{ac}}_{\mathrm{g}q}, \quad x^{\mathrm{dc}}_{\mathrm{fl}} = \Delta u^{\mathrm{dc}}_{\mathrm{g}} \end{aligned} \tag{22}$$

According to (4), the net input $\boldsymbol{e}_{\mathrm{c}}$ can be obtained via linear mapping of the above $\boldsymbol{x}_{\mathrm{c}}$, $\boldsymbol{x}_{\mathrm{fl}}$ and $\boldsymbol{u}_{\mathrm{ext}}=\begin{bmatrix}\Delta u^{\mathrm{ac}}_{\mathrm{g}d} & \Delta u^{\mathrm{ac}}_{\mathrm{g}q} & \Delta i^{\mathrm{dc}}_{\mathrm{g}}\end{bmatrix}^{\mathrm{T}}$, as in (23). To be consistent with (21), the actual variables are replaced with notations $\boldsymbol{x}_{\mathrm{c}}$, $\boldsymbol{x}_{\mathrm{fl}}$, $\boldsymbol{u}_{\mathrm{ext}}$ and original equations can be referred to Appendix A.

$$\begin{aligned} e_{\mathrm{c},1} &= \Delta u^{\mathrm{c}}_{\mathrm{g}q} = -s^{-1}U_{\mathrm{g}d0}x_{\mathrm{c},1} + u_{\mathrm{ext},2} \\ e_{\mathrm{c},2} &= -\Delta u^{\mathrm{dc}}_{\mathrm{g}} = -x^{\mathrm{dc}}_{\mathrm{fl}} \\ e_{\mathrm{c},3} &= \Delta Q = \frac{3}{2}\ U_{\mathrm{g}q0}x^{\mathrm{ac}}_{\mathrm{fl}d} - U_{\mathrm{g}d0}x^{\mathrm{ac}}_{\mathrm{fl}q} - I_{\mathrm{g}q0}u_{\mathrm{ext},1} + I_{\mathrm{g}d0}u_{\mathrm{ext},2} \\ e_{\mathrm{c},4} &= \Delta i^{\mathrm{c}*}_{\mathrm{g}d} - \Delta i^{\mathrm{c}}_{\mathrm{g}d} = -s^{-1}I_{\mathrm{g}q0}x_{\mathrm{c},1} + x_{\mathrm{c},2} - x^{\mathrm{ac}}_{\mathrm{fl}d} \\ e_{\mathrm{c},5} &= \Delta i^{\mathrm{c}*}_{\mathrm{g}q} - \Delta i^{\mathrm{c}}_{\mathrm{g}q} = s^{-1}I_{\mathrm{g}d0}x_{\mathrm{c},1} + x_{\mathrm{c},3} - x^{\mathrm{ac}}_{\mathrm{fl}q} \\ e_{\mathrm{c},6} &= \Delta u^{\mathrm{c}}_{\mathrm{g}d} = s^{-1}U_{\mathrm{g}q0}x_{\mathrm{c},1} + u_{\mathrm{ext},1} \\ e_{\mathrm{c},7} &= \Delta u^{\mathrm{c}}_{\mathrm{g}q} = -s^{-1}U_{\mathrm{g}d0}x_{\mathrm{c},1} + u_{\mathrm{ext},2} \end{aligned} \tag{23}$$

where $U_{\mathrm{g}d0}/U_{\mathrm{g}q0}$, $I_{\mathrm{g}d0}/I_{\mathrm{g}q0}$, $M_{d0}/M_{q0}$ are steady-state values of ac grid voltage, ac grid current, and modulation, respectively. Similarly, according to (9), the net input $\boldsymbol{e}_{\text{fl}}$ can also be obtained via linear mapping of $\boldsymbol{x}_{\text{c}}$, $\boldsymbol{x}_{\text{fl}}$ and $\boldsymbol{u}_{\text{ext}}$ as in (24), where $V_{\text{dc0}}/I_{\text{dc0}}$ are steady-states of dc voltage and dc current.

By assembling (23) with (24), matrix elements of (13) can be obtained as in (25), including six block matrices $\overline{\mathbf{A}}_{\text{c}}$, $\overline{\mathbf{B}}_{\text{c}}$, $\overline{\mathbf{C}}_{\text{c}}$, $\overline{\mathbf{A}}_{\text{fl}}$, $\overline{\mathbf{B}}_{\text{fl}}$, $\overline{\mathbf{C}}_{\text{fl}}$ divided by dashed lines. Then, by substituting ac/dc grid model (26) into (24) , the final feedback matrix $\boldsymbol{F}(s)$ as in (14) can be obtained, which is omitted here.

$$\begin{bmatrix} u_{\text{ext},1} \\ u_{\text{ext},2} \\ u_{\text{ext},3} \end{bmatrix} = \begin{bmatrix} sL_{\text{grid}} & \omega_1 L_{\text{grid}} & \\ \omega_1 L_{\text{grid}} & sL_{\text{grid}} & \\ & & -P_0\big/(V_{\text{dc0}})^2 \end{bmatrix} \begin{bmatrix} x_{\text{fl}d}^{\text{ac}} \\ x_{\text{fl}q}^{\text{ac}} \\ x_{\text{fl}}^{\text{dc}} \end{bmatrix} \tag{26}$$

*2) Model Validation via Frequency Scan*

Since $\boldsymbol{G}(s)$ of the GFC model is straightforward, in the next, only the feedback $\boldsymbol{F}(s)$ will be validated by frequency scan, the procedure of which is similar to the *dq* impedance measurement but with more perturbations to consider as the dimension of $\boldsymbol{F}(s)$ is larger. As given in Fig. 5(a), a series of perturbation signals $e_{\text{inj}}$ as shown in below will be injected once at a time;

$$e_{\text{inj}} = e_{\text{inj0}} \sin\left(\omega_{\text{inj}} t\right) \tag{27}$$

and, by collecting the responses from the input (left side of perturbation) and controller output of each channel, the measured $\boldsymbol{F}_{\text{mea}}(s)$ can be calculated as.

$$\boldsymbol{F}_{\text{mea}}\left(s\right) = \begin{bmatrix} e_{1,1} & \cdots & e_{1,n} \\ \vdots & \ddots & \vdots \\ e_{n,1} & \cdots & e_{n,n} \end{bmatrix} \begin{bmatrix} x_{1,1} & \cdots & x_{1,n} \\ \vdots & \ddots & \vdots \\ x_{n,1} & \cdots & x_{n,n} \end{bmatrix}^{-1} \tag{28}$$

where $n$ corresponds to the dimension of $\boldsymbol{F}(s)$ . Before implementing this method, it should be noted that the error inputs for $e_{\text{fl}d}^{\text{ac}}$, $e_{\text{fl}q}^{\text{ac}}$, $e_{\text{fl}}^{\text{dc}}$ physically correspond to the *dq*-voltage drop across the ac filter and the dc current of the dc capacitor, which are difficult to perturb in practice. To allow the frequency scan-based validation, the SISO equivalent concept [6] can be applied, i.e., apply *Kron*-reduction to the original model to find its equivalence of lower dimension. In this way, the three electric ports can be eliminated, and a port-reduced GFC model (i.e., $\boldsymbol{F}_{\text{RO}}$ ) can be derived according to (29).

$$\underbrace{\left[\begin{array}{c:c} \mathbf{0} & \\ \hdashline & \begin{pmatrix} Y_{\text{fl}dd} & Y_{\text{fl}dq} & \\ Y_{\text{fl}qd} & Y_{\text{fl}qq} & \\ & & Z_{\text{C}} \end{pmatrix}^{-1} \end{array}\right]}_{\boldsymbol{Y}} + \boldsymbol{F} = \left[\begin{array}{c:c} \left[\boldsymbol{Y}_{pp}\right]_{7\times7} & \left[\boldsymbol{Y}_{pn}\right]_{7\times3} \\ \hdashline \left[\boldsymbol{Y}_{np}\right]_{3\times7} & \left[\boldsymbol{Y}_{nn}\right]_{3\times3} \end{array}\right] \xrightarrow[\left[\boldsymbol{Y}_{pp}\right]_{7\times7}-\left[\boldsymbol{Y}_{pn}\right]_{7\times3}\left[\boldsymbol{Y}_{nn}^{-1}\right]_{3\times3}\left[\boldsymbol{Y}_{np}\right]_{3\times7}]{\text{apply Kron to find } \boldsymbol{Y}_{pp} \text{ equivalence, i.e., eliminate } \boldsymbol{Y}_{nn}} \boldsymbol{F}_{\text{RO}} \tag{29}$$

where $\boldsymbol{Y}_{nn}$ is a matrix with channels corresponding to the ac and dc filters. Then, this analytical model $\boldsymbol{F}_{\text{RO}}$ is compared with its frequency-scan result in Fig. 5(b), it can be seen that the measurements well coincide with the analytical curves, validating the effectiveness of the derived GFC model.

## *B. GFC Method Applied to the Multi-Converter Case*

*1) GFC Modeling*

For this case, a multi-converter study system is considered in Fig. 6(a), where two converters are homogenous in control structures for a clear demonstration. Then, based on (19) and (21), the forward matrix with explicit controller placement and clustering can be obtained as:

$$\begin{bmatrix} [x]_{\text{c},1} \\ [x]_{\text{c},2} \\ [x]_{\text{c},3} \\ [x]_{\text{c},4} \\ [x]_{\text{c},5} \\ [x]_{\text{c},6} \\ [x]_{\text{c},7} \\ \hdashline [x]_{\text{fl}d}^{\text{ac}} \\ [x]_{\text{fl}q}^{\text{ac}} \\ [x]_{\text{fl}}^{\text{dc}} \end{bmatrix} = \underbrace{\left[\begin{array}{ccccccc:ccc} [K]_{\text{pll}} &&&&&&&&& \\ & [K]_{\text{dc}} &&&&&&&& \\ && [K]_{\text{Q}} &&&&&&& \\ &&& [K]_{\text{cc}d} &&&&&& \\ &&&& [K]_{\text{cc}q} &&&&& \\ &&&&& [K]_{\text{u}d} &&&& \\ &&&&&& [K]_{\text{u}g} &&& \\ \hdashline &&&&&&& [Y]_{\text{fl}dd}^{\text{ac}} & [Y]_{\text{fl}dq}^{\text{ac}} & \\ &&&&&&& [Y]_{\text{fl}qd}^{\text{ac}} & [Y]_{\text{fl}qq}^{\text{ac}} & \\ &&&&&&&&& [Z]_{\text{fl}}^{\text{dc}} \end{array}\right]}_{\boldsymbol{G}(s)} \begin{bmatrix} [e]_{\text{c},1} \\ [e]_{\text{c},2} \\ [e]_{\text{c},3} \\ [e]_{\text{c},4} \\ [e]_{\text{c},5} \\ [e]_{\text{c},6} \\ [e]_{\text{c},7} \\ \hdashline [e]_{\text{fl}d}^{\text{ac}} \\ [e]_{\text{fl}q}^{\text{ac}} \\ [e]_{\text{fl}}^{\text{dc}} \end{bmatrix} \tag{30}$$

$$\begin{aligned} e_{\text{fl}d}^{\text{ac}} &= \Delta u_{\text{c}d}^{\text{ac}} - \Delta u_{\text{g}d}^{\text{ac}} = \left(\omega_1 L_{\text{fl}} I_{\text{g}d0} - V_{\text{dc0}} M_{q0}\right) s^{-1} x_{\text{c},1} + x_{\text{c},4} + x_{\text{c},6} - \omega_1 L_{\text{fl}} x_{\text{fl}q}^{\text{ac}} + M_{d0} x_{\text{fl}}^{\text{dc}} - u_{\text{ext},1} \\ e_{\text{fl}q}^{\text{ac}} &= \Delta u_{\text{c}q}^{\text{ac}} - \Delta u_{\text{g}q}^{\text{ac}} = \left(\omega_1 L_{\text{fl}} I_{\text{g}q0} + V_{\text{dc0}} M_{d0}\right) s^{-1} x_{\text{c},1} + x_{\text{c},5} + x_{\text{c},7} + \omega_1 L_{\text{fl}} x_{\text{fl}d}^{\text{ac}} + M_{q0} x_{\text{fl}}^{\text{dc}} - u_{\text{ext},2} \\ e_{\text{fl}}^{\text{dc}} &= \Delta i_{\text{c}}^{\text{dc}} - \Delta i_{\text{g}}^{\text{dc}} = \frac{3}{2V_{\text{dc0}}}\left(U_{\text{g}d0} x_{\text{fl}d}^{\text{ac}} + U_{\text{g}q0} x_{\text{fl}q}^{\text{ac}} + I_{\text{g}d0} u_{\text{ext},1} + I_{\text{g}q0} u_{\text{ext},2}\right) - \frac{I_{\text{dc0}}}{V_{\text{dc0}}} x_{\text{fl}}^{\text{dc}} - u_{\text{ext},3} \end{aligned} \tag{24}$$

$$\begin{bmatrix} e_{\text{c},1} \\ e_{\text{c},2} \\ e_{\text{c},3} \\ e_{\text{c},4} \\ e_{\text{c},5} \\ e_{\text{c},6} \\ e_{\text{c},7} \\ \hdashline e_{\text{fl}d}^{\text{ac}} \\ e_{\text{fl}q}^{\text{ac}} \\ e_{\text{fl}}^{\text{dc}} \end{bmatrix} = \left[\begin{array}{ccccccc:ccc:ccc} -s^{-1}U_{\text{g}d0} & 0 & 0 & 0 & 0 & 0 & 0 & 0 & 0 & 0 & 1 & 0 & 0 \\ 0 & 0 & 0 & 0 & 0 & 0 & 0 & 0 & 0 & -1 & 0 & 0 & 0 \\ 0 & 0 & 0 & 0 & 0 & 0 & 0 & 3U_{\text{g}q0}\big/2 & -3U_{\text{g}d0}\big/2 & 0 & -3I_{\text{g}q0}\big/2 & 3I_{\text{g}d0}\big/2 & 0 \\ -s^{-1}I_{\text{g}q0} & 1 & 0 & 0 & 0 & 0 & 0 & -1 & 0 & 0 & 0 & 0 & 0 \\ s^{-1}I_{\text{g}d0} & 0 & 1 & 0 & 0 & 0 & 0 & 0 & -1 & 0 & 0 & 0 & 0 \\ s^{-1}U_{\text{g}q0} & 0 & 0 & 0 & 0 & 0 & 0 & 0 & 0 & 0 & 1 & 0 & 0 \\ -s^{-1}U_{\text{g}d0} & 0 & 0 & 0 & 0 & 0 & 0 & 0 & 0 & 0 & 0 & 1 & 0 \\ \hdashline \omega_1 L_{\text{fl}} I_{\text{g}d0} s^{-1} - V_{\text{dc0}} M_{q0} s^{-1} & 0 & 0 & 1 & 0 & 1 & 0 & 0 & -\omega_1 L_{\text{fl}} & M_{d0} & -1 & 0 & 0 \\ \omega_1 L_{\text{fl}} I_{\text{g}q0} s^{-1} + V_{\text{dc0}} M_{d0} s^{-1} & 0 & 0 & 0 & 1 & 0 & 1 & \omega_1 L_{\text{fl}} & 0 & M_{q0} & 0 & -1 & 0 \\ 0 & 0 & 0 & 0 & 0 & 0 & 0 & 3U_{\text{g}d0}\big/2V_{\text{dc0}} & 3U_{\text{g}q0}\big/2V_{\text{dc0}} & -I_{\text{dc0}}\big/V_{\text{dc0}} & 3I_{\text{g}d0}\big/2V_{\text{dc0}} & 3I_{\text{g}q0}\big/2V_{\text{dc0}} & -1 \end{array}\right] \begin{bmatrix} x_{\text{c},1} \\ x_{\text{c},2} \\ x_{\text{c},3} \\ x_{\text{c},4} \\ x_{\text{c},5} \\ x_{\text{c},6} \\ x_{\text{c},7} \\ \hdashline x_{\text{fl}d}^{\text{ac}} \\ x_{\text{fl}q}^{\text{ac}} \\ x_{\text{fl}}^{\text{dc}} \\ \hdashline u_{\text{ext},1} \\ u_{\text{ext},2} \\ u_{\text{ext},3} \end{bmatrix} \tag{25}$$

where block matrices $[K]$, $[Y]$, $[Z]$ are the diagonal expansion forms of each matrix element in (21). Taking $[K]_{\text{pll}}$ and $[Y]_{\text{fl}dd}$ for example, they hold a structure as shown in (31).

$$[K]_{\text{pll}} = \begin{bmatrix} K[1]_{\text{pll}} & \\ & K[2]_{\text{pll}} \end{bmatrix}, \quad \boldsymbol{Y}_{\text{fl}dd} = \begin{bmatrix} Y[1]_{\text{fl}dd} & \\ & Y[2]_{\text{fl}dd} \end{bmatrix} \tag{31}$$

Based on (17) and (18), each element of the feedback matrix in (25) is expanded diagonally for the multi-converter system, e.g., the term $s^{-1}[I]_{gd0}$ in $\overline{\mathbf{A}}_{\text{c}}$, $\overline{\mathbf{B}}_{\text{fl}}$ should be expanded as:

$$s^{-1}[I]_{gd0} = \begin{bmatrix} s^{-1}I[1]_{gd0} & \\ & s^{-1}I[2]_{gd0} \end{bmatrix} \tag{32}$$

Then, the feedback matrix $\boldsymbol{F}(s)$ for multi-converter system can be obtained by the dimension-extended (25) and the ac/dc grids matrices (15). It is noted again that once the single converter model is obtained, $\boldsymbol{F}(s)$ can be acquired by directly vectorizing its elements if homogenous converter controls are considered.

*2) Validation via Marginal Stability Condition Test*

To validate the accuracy of the derived multi-converters' GFC model, marginal stability condition tests will be performed. This can be achieved by using the commonly-applied method with the impedance models, e.g., by calculating zeros of the determinant of $\boldsymbol{L}(s)$ from (33), and $\boldsymbol{L}(s)$ contains the forward and feedback gain of the GFC model for this work.

$$\boldsymbol{L}(s) = \boldsymbol{I} + \boldsymbol{G}(s)\boldsymbol{F}(s) \tag{33}$$

The mode distribution obtained from $\boldsymbol{L}(s)$ and under two types of grid SCR is shown in Fig. 6(b), one corresponds to a stable case (i.e., SCR =5) and the other is unstable (i.e., SCR =4.25). This theoretically-predicted stability results are then compared with time-domain simulations under the same conditions, the result is shown in Fig. 6(c). It can be seen that the simulated result shows the same stability trends as the theoretical prediction, particularly the oscillation frequency of theoretic prediction is almost the same as the time domain waveform (15 Hz $\approx 93/2\pi$). This analysis further confirms the validity of the GFC method for multi-converter systems.

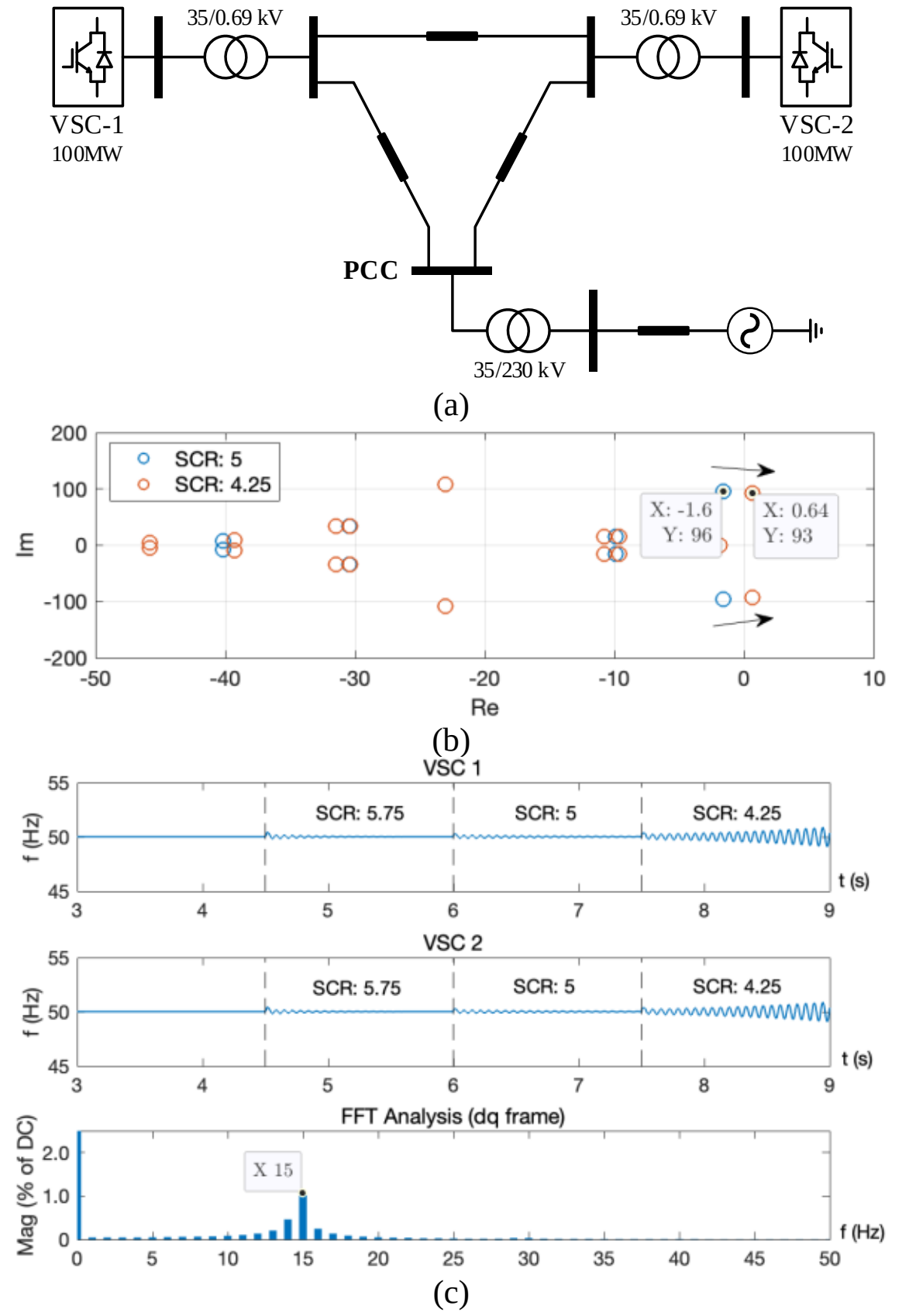

Fig. 6. GFC model validation for a multi converter system. (a) system topology; (b) mode distribution; (c) time-domain simulation validations.

## IV. Examples for Control-Driven Analysis

In this section, the capability and superiority of the GFC method in achieving better control-driven analysis and design will be demonstrated by three applications examples.

### *A. Qualitative Interaction Analysis of Multi-Controllers*

The GFC model enables a qualitative interaction analysis of different controller clusters given by its structure, this can be achieved by evaluating the frequency response of its open-loop gain, i.e., $\boldsymbol{G}(s)\boldsymbol{F}(s)$. To only focus on the interactions of actual controllers, the port-reduced GFC model is used (i.e., $\boldsymbol{Y}_{\text{fl}}^{\text{ac}}$ and $Z_{\text{fl}}^{\text{dc}}$ ports are eliminated as earlier applied). By substituting $s{=}j\omega$ into it, frequency responses of each matrix block element can be calculated as in Fig. 7, where controllers of the same type are merged together. Coordinate-*x*, -*y* are denoted as controller clusters and coordinate-*z* is the frequency. The color gradient is used to reflect the magnitude of response under each frequency, and the averaged magnitude response over the whole frequency range is quantified by the *width* of the 3D column bar. This means, the plot offers 5-dimensional information in total.

By inspecting Fig. 7, how distinct controller clusters are coupled and interacting at different frequencies can be perceived. For example, the non-ideal ac grid case in Fig. 7(a), one pair of coupling terms (i.e., upper and lower off-diagonal elements) exists between $K_{\text{pll}}$ and $K_{\text{cc}}$, $K_{\text{pll}}$ and $K_{\text{u}}$, indicating that PLL is simultaneously coupled with current inner loop and voltage feedforward. Whereas for the ideal ac grid case in Fig. 7(b), only upper off-diagonal elements exist, i.e., PLL is free from interacting with the other two control loops. This interpretation coincides with the common knowledge that PLL is more prone to interact with other loops in weak grids.

### *B. Frequency Domain Modal Analysis with GFC Model*

To better access the dominance of different types of controllers in closed-loop stability, the earlier mentiond FDMA method can be applied to the GFC model to extract the FD participation-factors (PF), i.e., the sensitivity of controller clusters to the FD eigenvalue of the model (i.e., eigen-loci). Worth noting that, since all controllers are explicitly placed in the forward matrix $\boldsymbol{G}(s)$ with GFC modeling, the calculation of their FD-PFs is straightforward, i.e., no need to apply the chain-rule of derivative as it does for the impedance-based system models [29]. To show how it works, the key steps are given below, while for more details about the calculation method, one may refer to, e.g., [24], [28].

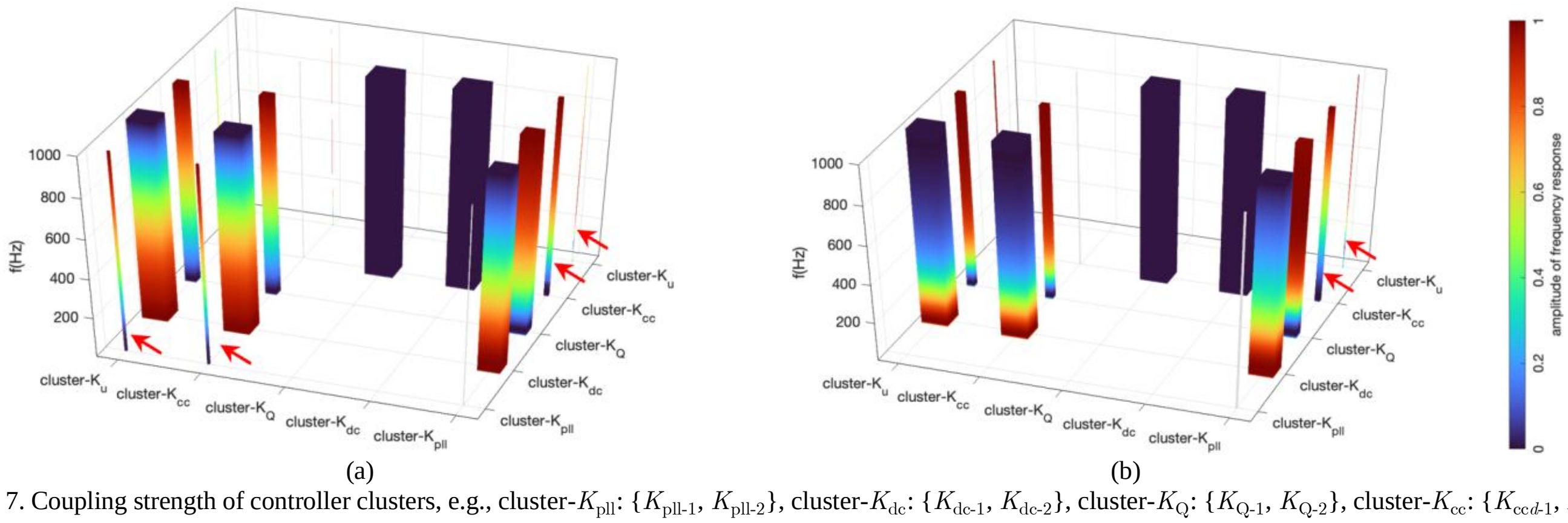

Fig. 7. Coupling strength of controller clusters, e.g., cluster-$K_{\mathrm{pll}}$: {$K_{\text{pll-1}}$, $K_{\text{pll-2}}$}, cluster-$K_{\mathrm{dc}}$: {$K_{\text{dc-1}}$, $K_{\text{dc-2}}$}, cluster-$K_{\mathrm{Q}}$: {$K_{\text{Q-1}}$, $K_{\text{Q-2}}$}, cluster-$K_{\mathrm{cc}}$: {$K_{\text{cc}d\text{-1}}$, $K_{\text{cc}q\text{-1}}$, $K_{\text{cc}d\text{-2}}$, $K_{\text{cc}q\text{-2}}$}, cluster-$K_{\mathrm{u}}$: {$K_{\text{u}d\text{-1}}$, $K_{\text{u}q\text{-1}}$, $K_{\text{u}d\text{-2}}$, $K_{\text{u}q\text{-2}}$}. (a) non-ideal ac grid with SCR=5; (b) ideal ac grid ignoring impedances of ac lines and transformers.

The first step is to apply the eigen-decomposition to $\boldsymbol{L}(s)$, as:

$$\Lambda_k = \boldsymbol{t}_k \boldsymbol{L} \boldsymbol{r}_k = \boldsymbol{t}_k \left( \boldsymbol{I} + \boldsymbol{G}\boldsymbol{F} \right) \boldsymbol{r}_k \tag{34}$$

where $\Lambda_k$ is the $k$-th FD-eigenvalue [28], $\boldsymbol{r}_k$ and $\boldsymbol{t}_k$ are the associated right and left eigenvectors. Then, the sensitivity of $\Lambda_k$ to the controller $G_{ii}$ can be calculated as:

$$\frac{\partial \Lambda_k}{\partial G_{ii}} = \boldsymbol{t}_k \left( \frac{\partial \boldsymbol{G}}{\partial G_{ii}} \boldsymbol{F} \right) \boldsymbol{r}_k = t_{ki} \boldsymbol{F}_{i\text{-row}} \boldsymbol{r}_k \tag{35}$$

Collecting the eigen-sensitivities of all the controllers, yields:

$$\left[ \frac{\partial \Lambda_k}{\partial G_{11}} \cdots \frac{\partial \Lambda_k}{\partial G_{nn}} \right] = \left[ t_{k1} \boldsymbol{F}_{1\text{-row}} \boldsymbol{r}_k \ \cdots \ t_{kn} \boldsymbol{F}_{n\text{-row}} \boldsymbol{r}_k \right] \tag{36}$$

For the studied case in Fig. 6(a), two FD-eigenvalues $\Lambda_{10}$ and $\Lambda_{13}$ with the presence of significant resonance points (i.e., $\Lambda \approx 0$, 4Hz for $\Lambda_{10}$, and 15Hz for $\Lambda_{13}$) will be focused, as shown in Fig. 8. Then, the FD-PF of each controller cluster is calculated based on (36). The resulting curves can be compared under the shaded area to acquire information on the dominant controller clusters. E.g., first, the 4Hz resonance area in Fig. 8(a) is analyzed, it can be seen that under this area cluster-$K_{\mathrm{dc}}$ and cluster-$K_{\mathrm{cc}}$ are of the most significant influence due to their relatively large values. As for the 15Hz resonance area in Fig. 8(b), cluster-$K_{\mathrm{pll}}$ and cluster-$K_{\mathrm{cc}}$ becomes dominant. This analysis demonstrates that the GFC model can empower the traditional FD-PF method by enabling a more efficient acquisition of controllers' sensitivities.

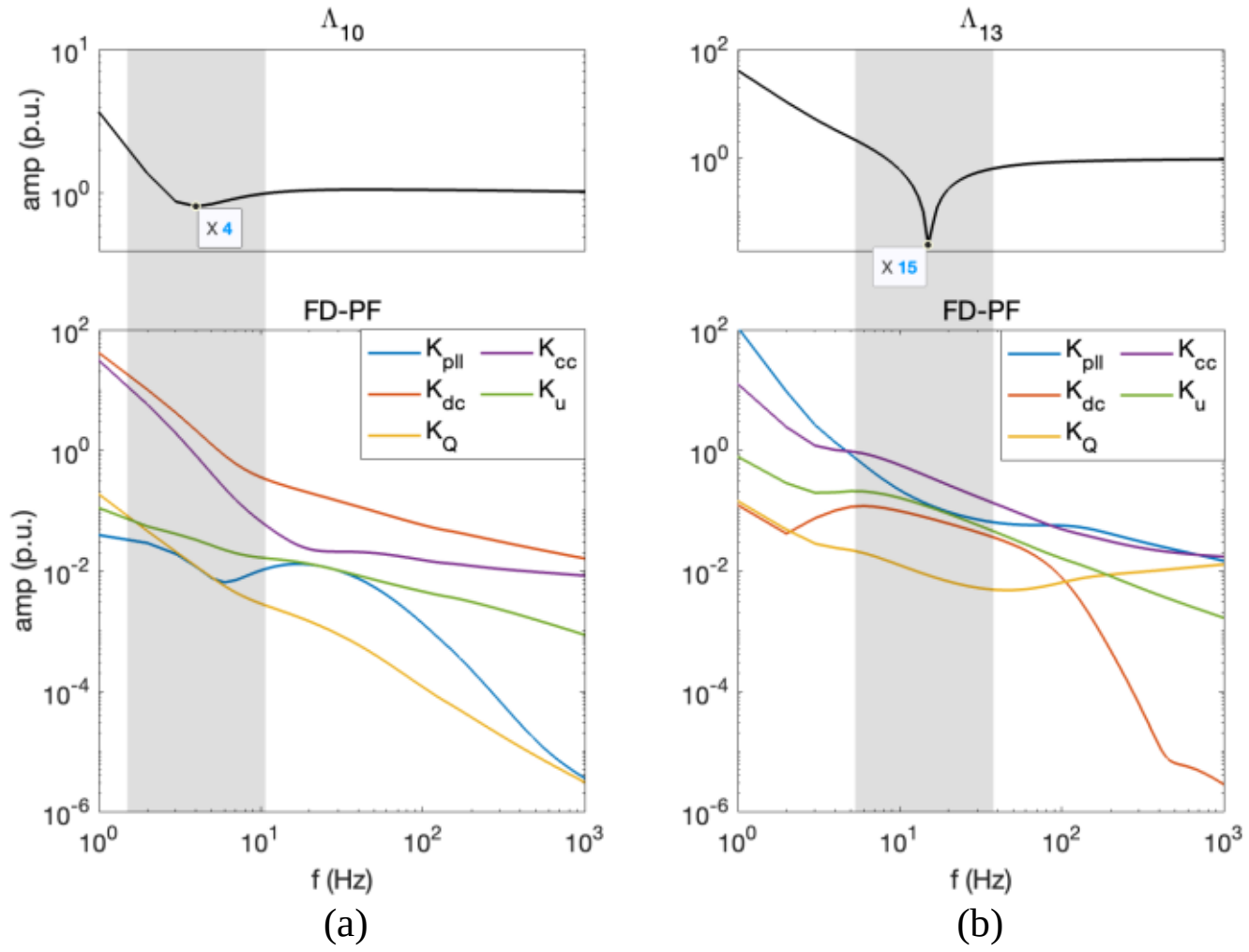

Fig. 8. FD-PF of various controller clusters. For concise analysis, the PF matrix for each controller cluster with identical controllers is evaluated by 2-norm.

## C. Stability-oriented Multi-Controller Reshaping

In this part, a simple example in achieving stability-oriented multi-controllers reshaping with the GFC model is provided. This is primarily achieved by adding an ancillary feedback loop $\gamma_i(s)$ to shape the behavior of the original controllers, thereby system stability is improved, as shown in Fig. 9 (a).

To quantitatively design the stability issue, time-domain (TD) eigenvalue sensitivity is desired. In this respect, [24], [28] proposed a translator for TD and FD eigen-sensitivity, enabling the acquisition of TD eigen-sensitivity by using the FD models (e.g., GFC model). With this information, the desired increments of parameters for a specified mode-movement can be quantified, which is formulated as an optimization routine for this case. Main steps to arrive at this is shown below.

The first step is to transform the FD-PF into TD-PF.

$$\frac{\partial \lambda_m}{\partial G_{ii}} = \xi_{km} \frac{\partial \Lambda_k}{\partial G_{ii}} \tag{37}$$

where $\Lambda_k$ is FD-eigenvalue, $\lambda_m$ denotes a "weak" or intended mode $\lambda_m$ for improvements, $\xi_{km}$ is the TD-FD translator [28].

Then, the difference of $\lambda_m$ can be approximated by first-order derivatives of multiple controllers as (38), of which the TD partial derivatives can be finally calculated by FD-sensitivities according to (37).

$$\Delta \lambda_m = \frac{\partial \lambda_m}{\partial G_{11}} \Delta G_{11}(\lambda_{m0}) + \cdots \frac{\partial \lambda_m}{\partial G_{nn}} \Delta G_{nn}(\lambda_{m0}) \tag{38}$$

where $G_{ii}$ is an element of $\boldsymbol{G}(s)$, i.e., controller.

Afterward, the change of controller due to the adding of the ancillary control loop $\gamma_i(s)$ can be further formulated as:

$$\Delta G_{ii}(\lambda_{m0}) = \frac{K_i(\lambda_{m0})}{1 + K_i(\lambda_{m0})\gamma_i(\lambda_{m0})} - K_i(\lambda_{m0}) \tag{39}$$

where $\gamma_i(s)$ is considered to be low- or high-pass filter as:

$$\gamma_i(s) = D_i \frac{\omega_{\mathrm{c}i}}{s + \omega_{\mathrm{c}i}} \quad or \quad \gamma_i(s) = D_i \frac{s}{s + \omega_{\mathrm{c}i}} \tag{40}$$

with two unknowns under determined or tuned, i.e., $D$ and $\omega_{\mathrm{c}}$.

Finally, the stability-oriented optimization routine for multi-parameters tuning of multi-controllers is formulated as (41), where $\Delta\lambda_m^*$ denotes the desired incremental movement of the mode, which is usually set as a negative value for stability improvement (i.e., moving the mode to the left).

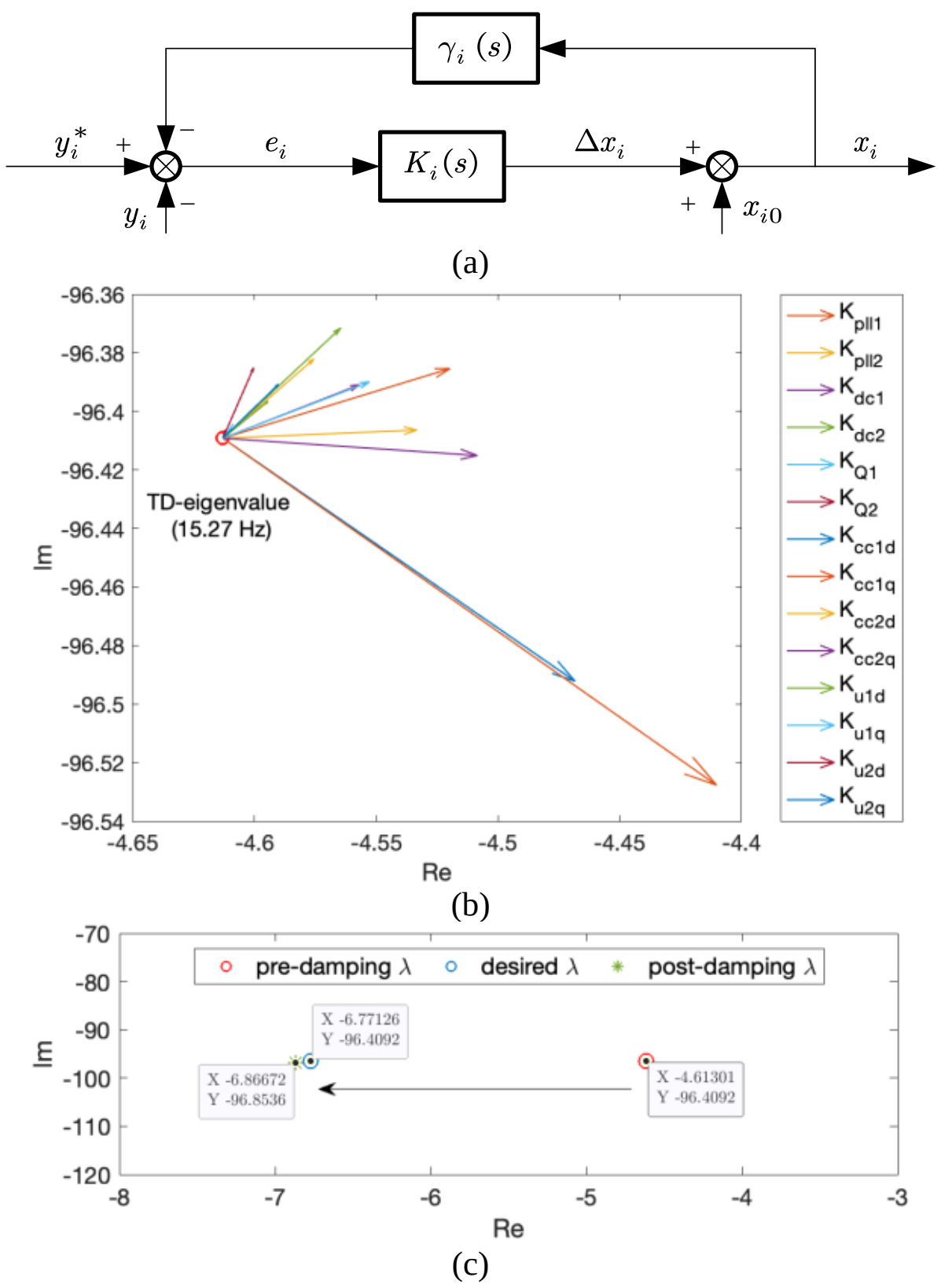

Fig. 9. Stability-oriented reshaping of multiple controllers. (a) structure for reshaping controller *i* by adding an ancillary loop; (b) FD-PF of controllers with respect to the TD-eigenvalue; (c) mode position before and after reshaping.

$$
min: \left|\Delta\lambda_m^* - \Delta\lambda_m\right|
$$

$$
subject\ to: \begin{Bmatrix} \text{Eq. (38),with (37) and (39)} \\ D_{\min} < D_i \le D_{\max} \\ \omega_{\min} < \omega_{ci} \le \omega_{\max} \end{Bmatrix},\quad 1 \le i \le n \tag{41}
$$

Then, by solving this problem, the desired parameters of $\omega_i$ and $D_i$ of all the controllers can be obtained at a time.

The results of this analysis and its validation are given in Fig. 9(b), (c) and Table I. First, the calculated TD-PF information of each controller is shown in Fig. 9(b), from which effect of the controllers on the movement of mode can be acquired. Then, a desired mode position is input into (41) and all the underdetermined parameters from the ancillary loops are solved simultaneously as shown in Table I. With the above optimized parameters, the weak mode (pre-damping) is expected to move by $\Delta\lambda_m^*$. To validate this, the loop gain model $\boldsymbol{L}(s)$ is updated with the calculated ancillary loop configurations, and the mode is recalculated as in Fig. 9(c). It can be seen that after parameter optimization, the system mode settles down at a place adequately close to the reference mode, indicating the stability margin is improved.

TABLE I
RESULTS OF MULTI-CONTROLLERS' DAMPING DESIGN

| Controllers | Damping method | $D$ (p.u.) | $\omega_c$ (Hz) |
|---|---|---|---|
| $K_{\text{pll-1}}(s)$ | high pass filter | 3.1607 | 42.2476 |
| $K_{\text{pll-2}}(s)$ | high pass filter | 5.5235 | 1.8904 |
| $K_{\text{dc-1}}(s)$ | low pass filter | 0.0000 | 0.3116 |
| $K_{\text{dc-2}}(s)$ | low pass filter | 0.0023 | 0.0349 |
| $K_{\text{Q-1}}(s)$ | low pass filter | 0.2777 | 0.0031 |
| $K_{\text{Q-2}}(s)$ | low pass filter | 0.1117 | 0.0033 |
| $K_{\text{cc}d\text{-1}}(s)$ | low pass filter | 9.9815 | 1.1620 |
| $K_{\text{cc}q\text{-1}}(s)$ | low pass filter | 9.9110 | 0.1710 |
| $K_{\text{cc}d\text{-2}}(s)$ | low pass filter | 0.0892 | 0.7417 |
| $K_{\text{cc}q\text{-2}}(s)$ | low pass filter | 0.5484 | 0.1822 |
| $K_{\text{u}d\text{-1}}(s)$ | low pass filter | 5.0929 | 1.5081 |
| $K_{\text{u}q\text{-1}}(s)$ | low pass filter | 0.0000 | 65.0422 |
| $K_{\text{u}d\text{-2}}(s)$ | low pass filter | 0.0499 | 84.0692 |
| $K_{\text{u}q\text{-2}}(s)$ | low pass filter | 0.0002 | 13.7146 |

Overall, the merit of the GFC model from both the "control" and "analysis" aspects are demonstrated by the above examples.

## V. CONCLUSIONS

To cope with the issues of control-driven analysis and design of converter-dominated systems with large-scale controllers, a new frequency-domain modeling framework and method is proposed, referred to as the GFC modeling method.

The effectiveness and accuracy of the proposed modeling method are validated by both single and multi-converter systems. And, the merits of the GFC model are demonstrated by the qualitative interaction analysis of controller clusters and the quantitative multi-controller reshaping designs for stability improvements, in which the frequency-domain modal analysis (FDMA) method that usually used with the impedance models is applied with the GFC model to show its good compatibility to existing frequency-domain tools.

Overall, the GFC modeling framework due to its explicit controller placement and controller clustering trait addressed the deficiency of the impedance-based methods in tuning and designing large-scale controllers. Moreover, it in conjunction with the existing FD tools, shows a prospective tool for control-intensive converter systems analysis and design.

### *APPENDIX A*

Other converter's *dq*-frame equations are given in (42), i.e., the frame transformation①, ②, ⑤, the power calculation ③, the modulation ④, ⑥, and the ac-dc power balance ⑦.

$$
\begin{aligned}
&① : \left[\Delta u_{gd}^{c} \quad \Delta u_{gq}^{c}\right]^{\mathrm{T}} = \left[\Delta u_{gd}^{ac} \quad \Delta u_{gq}^{ac}\right]^{\mathrm{T}} + s^{-1}\left[U_{gq0} \quad -U_{gd0}\right]^{\mathrm{T}} \Delta\omega \\
&② : \left[\Delta i_{gd}^{c} \quad \Delta i_{gq}^{c}\right]^{\mathrm{T}} = \left[\Delta i_{gd}^{ac} \quad \Delta i_{gq}^{ac}\right]^{\mathrm{T}} + s^{-1}\left[I_{gq0} \quad -I_{gd0}\right]^{\mathrm{T}} \Delta\omega \\
&③ : \begin{bmatrix}\Delta P \\ \Delta Q\end{bmatrix} = \frac{3}{2}\begin{bmatrix}U_{gd0} & U_{gq0} \\ U_{gq0} & -U_{gd0}\end{bmatrix}\begin{bmatrix}\Delta i_{gd}^{ac} \\ \Delta i_{gq}^{ac}\end{bmatrix} + \frac{3}{2}\begin{bmatrix}I_{gd0} & I_{gq0} \\ -I_{gq0} & I_{gd0}\end{bmatrix}\begin{bmatrix}\Delta u_{gd}^{ac} \\ \Delta u_{gq}^{ac}\end{bmatrix} \\
&④ : \left[\Delta u_{cd}^{ac} \quad \Delta u_{cq}^{ac}\right]^{\mathrm{T}} = \left[\Delta m_d \quad \Delta m_q\right]^{\mathrm{T}} V_{\mathrm{dc0}} + \left[M_{d0} \quad M_{q0}\right]^{\mathrm{T}} \Delta u_{\mathrm{g}}^{\mathrm{dc}} \\
&⑤ : \left[\Delta m_d^{c} \quad \Delta m_q^{c}\right]^{\mathrm{T}} = \left[\Delta m_d \quad \Delta m_q\right]^{\mathrm{T}} + s^{-1}\left[M_{q0} \quad -M_{d0}\right]^{\mathrm{T}} \Delta\omega \\
&⑥ : \begin{bmatrix}\Delta m_d^{c} \\ \Delta m_q^{c}\end{bmatrix} = \frac{1}{V_{\mathrm{dc0}}}\left(\begin{bmatrix}\Delta u_{\mathrm{cd1}}^{c} \\ \Delta u_{\mathrm{cq1}}^{c}\end{bmatrix} + \begin{bmatrix}\Delta u_{\mathrm{cd2}}^{c} \\ \Delta u_{\mathrm{cq2}}^{c}\end{bmatrix} + \begin{bmatrix}0 & -\omega_1 L_{\mathrm{fl}} \\ \omega_1 L_{\mathrm{fl}} & 0\end{bmatrix}\begin{bmatrix}\Delta i_{gd}^{c} \\ \Delta i_{gq}^{c}\end{bmatrix}\right) \\
&⑦ : \Delta i_{\mathrm{c}}^{\mathrm{dc}} = V_{\mathrm{dc0}}^{-1}\left(\Delta P - \Delta u_{\mathrm{g}}^{\mathrm{dc}} I_{\mathrm{dc0}}\right)
\end{aligned} \tag{42}
$$